\documentstyle[12pt]{article}
\def\doublespacing {\parskip 5 pt plus 1 pt \baselineskip 25 pt
                    \lineskip 13 pt \normallineskip 13 pt}
\newcommand{\bold}[1]{\mbox{\boldmath $#1$}}    
\begin{document}
\begin{titlepage}

\begin{center}
{\large\bf Coexistence of Regular Undamped Nuclear
           Dynamics with Intrinsic Chaoticity}\\[4ex]

{\sl Wolfgang Bauer, Vladimir Zelevinsky
}\\[2ex]

National Superconducting Cyclotron Laboratory\\
and Department of Physics and Astronomy\\
Michigan State University\\
East Lansing, Michigan 48824-1321, USA\\[4ex]

and {\sl Peter Schuck}\\[2ex]

Institut de Physique Nucl\'eaire\\
Universit\'e de Grenoble\\
53 avenue des Martyrs, 38026 Grenoble Cedex, France\\[8ex]

\end{center}

\begin{abstract}
We study the conditions under which the nucleons inside a deformed nucleus can
undergo chaotic motion.  To do this we perform self-consistent calculations in
semiclassical approximation utilizing a multipole-multipole interaction of the
Bohr-Mottelson type for quadrupole and octupole deformations. For the case of
harmonic and non-harmonic static potentials, we find that both multipole
deformations lead to regular motion of the collective coordinate, the multipole
moment of deformation.  However, despite this regular collective motion, we
observe chaotic single particle dynamics.

\end{abstract}

\vspace*{1cm}
\centerline{{\bf PACS}: 21.10.Ky, 21.10.Re, {\bf 21.60.-n}, 24.60.Lz}

\end{titlepage}

\doublespacing

\newpage

The question about the origin of dissipation in collective motion of finite
Fermi systems \cite{HW53}
such as atomic nuclei or small metallic clusters is an intriguing
and up to now not completely satisfactorily solved problem.  For example,
the mutual balance of one-body and two-body processes is still a question of
debate.  For the case of one-body dissipation and friction in nuclear
dynamics, Swiatecki and coworkers \cite{BBN78,RS80,BBS92,BS93}
have developed this picture:
particles which move in a shape-deformed container are reflected from
the (moving) walls, and due to parts of it having positive curvature (for
higher multipole moments) the particles very quickly loose their
coherence, thus inducing pseudo-random motion, i.e.\ heat, into the system.
At the same time the shape oscillation is very much slowed down.

Blocki {\it et al.}\ \cite{BBS92,BS93} consider a purely classical
gas of particles contained in a deformed billiard.  The only similarity
with a Fermi gas comes from the fact that initially the particles'
momenta are distributed within a Fermi sphere.  The walls of the
container undergo periodic shape oscillations with a frequency much
smaller than a typical single particle frequency.  In the
interior of the container the particles move on linear trajectories.
They study the particle kinetic
energy increase as a function of time and find that for ellipsoidal
shape deformations ($\ell=2$) the particles act as a classical Knudsen
gas \cite{Knu50}, i.e.\ the total kinetic energy increase over an entire shape
oscillation period is 0.
However, for $\ell\ge 3$ the kinetic energy in the single particle motion
is not completely `given back', but rather steadily increases in time.
This is explained by the fact that in an $\ell=2$ potential the motion
of the particles remains non-chaotic and therefore coherent, whereas in
the $\ell\ge 3$ the scattering of the segments of the wall with
positive curvature leads to chaotic motion similar to the one observed
in a Sinai \cite{Sin70,BB90} billiard and thus a destruction of
coherence.  The fact that deformed nuclear potentials may exhibit chaotic
motion was recognized early by Arvieu and co-workers \cite{Arv}

This scenario is very similar to
the so-called Fermi acceleration, proposed to explain the occurrence of
very high energy cosmic radiation \cite{Fer49,LL83}.

In this paper we present an attempt to include
\underline{selfconsistency} into the problem
of motion in multipole-deformed nuclear potentials.
We have chosen a selfconsistent, but schematic, model
of separable forces.  We chose an interaction of the Bohr-Mottelson type
\cite{BM75} with a static $r^2$ potential and multipole-multipole
interactions as studied, for example, by Stringari {\it et al.}
\cite{Str79,Str81,RS82,KSS86}.
In the small amplitude limit, this model has recently been investigated
in the semiclassical limit \cite{KSS86}; as is known, the
low-lying quadrupole and
octupole frequencies come out to be in reasonable
agreement with experimental data.
Our single-particle Hamiltonian is then
\begin{eqnarray}
  {\cal H} &=& {\cal H}_0 + V^{(\ell)}(\bold{r},t) \nonumber \\
           &=& \frac{p^2}{2\,m} + V_0 + V^{(\ell)}(\bold{r},t)
\label{eq2}
\end{eqnarray}
$V^{(\ell)}(\bold{r},t)$ is the potential associated with the
(separable) multipole-multipole force \cite{Str79,Str81,KSS86}
\begin{equation}
\label{eqVell}
  V^{(\ell)}(\bold{r},t) = \lambda_\ell\,q_\ell(\bold{r})\,Q_\ell(t)\ ,
\end{equation}
and $V_0$ is the static external potential.  We take here
$V_0 = \frac{1}{2}\,m\,\omega_0^2\,r^2$, resulting in the Bohr-Mottelson
Hamiltonian \cite{BM75}.  However, we have also investigated non-harmonic
static potentials ($r^6$) and obtained similar results.

The coupling constants $\lambda_\ell$ can be
calculated using a self-consistent normalization condition
\cite{BM75,KSS86}.
\begin{eqnarray}
  \lambda_{2,s} &=& - \frac{3\,m\,\omega_0^2}{A\langle r^2\rangle}\ ,\\
  \lambda_{3,s} &=& - \frac{15\,m\,\omega_0^2}{A\langle r^4\rangle}\ ,
\end{eqnarray}
where $A$ is the mass number of the nucleus under consideration.
$q_\ell(\bold{r})$ is given by
\begin{eqnarray}
  q_2(\bold{r}) &=& r_y\,r_z\\
  q_3(\bold{r}) &=& r_x\,r_y\,r_z
\end{eqnarray}
and the multipole moments $Q_\ell(t)$ are
\begin{equation}
  Q_\ell(t) = \int \frac{d^3r\,d^3p}{(2\pi)^3} q_\ell(\bold{r})
                   f(\bold{r},\bold{p},t)\ .
\end{equation}
$f(\bold{r},\bold{p},t)$ is the one-body phase space
distribution function of nucleons, the Wigner transform of the
one-body density.

We treat this problem in semi-classical approximation by
a Wigner transformation
of the von Neumann equation of motion for the density matrix,
$i\,\partial_t\rho = [{\cal H},\rho]$, to obtain a
Vlasov equation, $\partial_t f = \{{\cal H},f\}$.
We then solve the Vlasov equation in the test particle method
\cite{MB82,Won82} using a fourth-order Runge-Kutta algorithm with typical
time step sizes of 1 fm/c.
Our numerical simulation is fully selfconsistent and conserves total energy
to better than 0.1 \%.

In order to generate selfconsistent initial deformations in coordinate and
momentum space, we start with a spherically symmetric configuration generated
in local Thomas-Fermi approximation without the deformation potential
$V^{(\ell)}$.

We now apply a time-dependent external potential of the form
\begin{equation}
\label{V_ext}
  V^{(\kappa)}_0(\ell,\bold{r},t) = \kappa_0\,\sin(\omega_D\,t)\,s(t)
     \,q_\ell(\bold{r})
\end{equation}
where $\omega_D$ is the driving frequency, and
where $s(t)$ is a differentiable spline interpolation function on the time
interval [0,$\tau$] with vanishing first derivatives at both ends, which is
monotonically increasing from 0 to 1.
This procedure results in a giant oscillation of the nucleus at $t=\tau$,
provided that $\tau$ is chosen $\tau\gg\omega_D^{-1}$.  The deformation
is dependent on the value of the coupling $\kappa_0$ chosen.

We now use the initial conditions generated in this way (with $\tau$ = 1500
fm/c) to study the time evolution
under the action of our Hamiltonian as defined in Eq.\ (\ref{eq2}).
The upper panel (a) of Figure 1 contains the results
of our calculations.  One can clearly observe a regular undamped
oscillation of the quadrupole moment in coordinate space as a function
of time.  One can observe that the period of oscillation has been
stretched from the 0-coupling value of 88.4 fm/c to 128 fm/c.  This is
consistent with the analytic calculations for infinitesimal
deformations \cite{Suz73,BM75,KSS86}, which yield
\begin{eqnarray}
  Q_2(t) &=& Q_2(t_0)\,\exp(i\,\omega_{2^+}\,t)\nonumber\\
  \omega_{2^+} &=& \sqrt{4\omega_0^2 + \lambda_2\frac{2\,A\,\langle
                   r^2\rangle}{3m}}\nonumber\\
               &=& \sqrt{2}\,\omega_0
\end{eqnarray}
for the giant quadrupole frequency, and consequently $T=125$ fm/c for
$\omega_0 = 0.0355$ c/fm used in this example.

In the lower panel (b) of Figure 1 we show our calculations for the
octupole case.
Again we see harmonic oscillations (The small variations in amplitude
are in both cases due to beating between the initial driving frequency
and the oscillation frequency of the self-consistent calculation.).
The observed oscillation period is $T_{3} \approx 66$ fm/c, in good
agreement with the analytical result of \cite{KSS86}
\begin{equation}
  \omega_{3^-} = \sqrt{7}\,\omega_0
\end{equation}
for the giant octupole oscillation frequency, which results in
$T_{3} \approx 66.8$ fm/c for our value of $\omega_0$.

The most important observation is here, however,
that there is no damping of the collective
motion apparent in our calculation, thus indicating that no
chaoticity is present in the collective multipole coordinates.
We have also used slightly
different initial conditions (by using a different number of test particles
in the simulation) and obtained only slightly different
results.  This indicates that there is no sensitive dependence on the
initial conditions present here as would be the case for chaotic
motion.  As an additional test, we performed a Fourier transform of
the time signal and found one peak at the dominant frequency and no
$\omega^{-1}$ noise.

To study the single particle dynamics we employ the method of Poincar\'e
surface of sections.
In Poincar\'e sections, we stroboscopically record the phase space coordinates
of particles in time.  This `stroboscope' is triggered whenever the particles
cross a certain plane in phase space.
This technique is also applicable to our problem, where
we solve the Vlasov equation utilizing the test particle method.
Here we choose the plane $r_x = 0$ as the trigger condition.

In Figure 2 (a) we show the Poincar\'e section for one test particle used
to generate the octupole motion of Fig.\ 1 (b).  One can clearly see that
this Poincar\'e section will become area-filling in certain regions of phase
space, and in the limit of very long
time scales. (For the present calculations we ran 10$^5$ time steps of 1 fm/c
each, leading to approximately 550 crossings of the $r_x=0$ plane from
below).
In order to enable the reader to better compare the structures, we superimpose
in Figure 2 (b) the Poincar\'e sections of ten test particles
with different initial conditions (but parts of the same simulation).
We obtain similar results for the collective quadrupole motion.

This chaoticity is not a result of the weak destruction of the
integrability of the corresponding static multipole
potential, which can be shown to exist for the static octupole (but not
for the static quadrupole) potential.
Instead we attribute the chaoticity in the single particle motion
to the exchange of energy between the motion of the individual test particles
and the collective motion of the multipole coordinate. This
exchange of energy is possible, because the individual test
particles oscillate with frequencies, which do not have a rational ratio
with the frequency of the
collective coordinate.   This results in the
particle reaching meta-stable or unstable points in phase space during the
course of its time evolution.  At these points small changes in the initial
conditions will have a large effect on the subsequent dynamics.  An example
for this would be the decision if the particle will temporarily
oscillate in or out of phase with the collective coordinate.  Consequently,
these points provide large positive contributions to the Kolmogorov entropy,
and chaotic single particle dynamics results.

In turn one also expects each single test particle to
have a randomly fluctuating effect on the energy contained in the motion
of the collective coordinate.  However, since there are quite many test
particles, these chaotic random fluctuations are averaged out leaving only
a smooth sinusoidal oscillation of the collective coordinate.
This is qualitatively new in our investigation: the generation of regular
dynamics for the collective variable, the multipole moment of the collective
oscillation, from the ensemble of single particles with chaotic trajectories.
This is an example of how ordered macroscopic motion can result form underlying
chaotic microscopic dynamics.  (To obtain this result, it was crucial
to employ a self-consistent treatment of the dynamics entailing conservation
of total energy.)

Our finding are not restricted to the static harmonic potential, which we used
as an example here, but they hold for a general class of
central potentials, $V_0$.
We have also performed calculations for a $V_0 \propto r^6$ potential
with very similar results to the ones presented here \cite{More}.

And, finally, one may speculate that this interplay between chaoticity in
individual single particle degrees of freedom and regularity in certain
collective coordinates may also play a role in the time evolution of other
physical systems.  Examples that come to mind as likely candidates are
plamas in a tokamak, the human brain wave activity, the weather.  Chaos on
a microscopic level need not necessarily lead to a catastrophic breakdown
of the system on the macroscopic scale.

This work was supported in parts by the US National Science Foundation under
grant number PHY 90-17077.  W.B. acknowledges support form a US
NSF Presidential Faculty Fellow award.  P.S. is grateful for useful discussions
with J. Blocki.

\clearpage

\section*{Figure Captions}

\begin{description}

\item[Figure 1]\hspace*{1cm}\\
      Time evolution of quadrupole (a) and octupole (b) deformations of an
      $A$=200 nucleus in
      coordinate space under propagation with the Hamiltonian of Eq.\
      (\ref{eq2}) with a static harmonic oscillator potential and
      selfconsistent coupling strength
      $\lambda_2 = \lambda_{2,s} = 8.9\cdot 10^{-4}$ MeV\,fm$^{-4}$ (a),
      and $\lambda_3 = \lambda_{3,s} = 1.3\cdot 10^{-4}$ MeV\,fm$^{-6}$
      (b).

\item[Figure 2]\hspace*{1cm}\\
      Poincar\'e section of a single test particle (a) and an ensemble of
      10 test particles (b) from the calculation leading to the time evolution
      of the octupole moment shown in Fig.\ 1 (b).

\end{description}


\begin{thebibliography}{99}
\bibitem{More}  A more complete study including comparison to the `wall
                formula' results as well as a presentation of the Fourier
                spectra and the discussion of the destruction of integrability
                in the static octupole potential will be published elsewhere.
\end{thebibliography}
\end{document}